\documentclass[11pt]{article}
\usepackage{epsfig}
\def \beq{\begin{equation}}
\def \eeq{\end{equation}}
\begin{document}
\author{ {\bf Jerzy Szwed}\footnote {Work supported in part by the Marie Curie Actions Transfer of
Knowledge project COCOS (contract MTKD-CT-2004-517186).}
\\
Institute of Physics, Jagellonian University,
\\ 
Reymonta 4, 30-059 Krak\'ow, Poland
}
\title{\Large \bf Interference Effects in the Brain.}
\maketitle

TPJU - 1/2008

\begin{abstract}
Interference effects are the most spectacular manifestation of the wave nature of phenomena. 
This note proposes a systematic search for such effects in the brain. 
\end{abstract}

1. The current discussion on possible new effects playing an important role in neural processes in the brain is provoking many speculative hypotheses. 
Among them the suggestion of nonalgorithmic processes taking place in the mind or the quantum nature of information processing is perhaps
 the most exciting \cite{Penrose}. 
In our opinion further theoretical development of these ideas requires definite feedback from experiment. In this note we propose a series 
of experiments testing the possibility of wave phenomena taking place and playing an important role in the brain. 
Inspiration for these proposals comes from analogous experiments in physics.

2. The fundamental experiment proving the wave nature of light was performed by Thomas Young in 1801 \cite{Young}. 
Although well known to physicists we describe it briefly
because of its central role in further considerations. Put simply, the experimental setup consisted  of a (nearly) pointlike 
light source  and two slits between the source and screen (see Fig.~1). Covering one of the two slits  
(Fig.~1a) the light travelled through the second slit 
and hit the screen producing a (slightly diffused) picture opposite the opened slit. Today we know that what we observe is the Huygens principle at work --- 
the open slit works as the source of a spherical wave. Analogically, covering the second slit we observe the picture on the screen opposite the first slit (Fig.~1b). 
With both slits opened a spectacular result is observed (Fig.~1c). It is not the simple sum of both previous pictures, instead an interference pattern of dark and light bands
appears. Moreover, at maximum brightness, the signal is not just twice as strong (as one would expect from summing the two separate signals) but 4 times stronger. 
Today the explanation is well known --- what is travelling between the source and screen are the light (electromagnetic) waves. On the screen these waves meet 
having covered in general different distance when travelling through the first and the second slit. If the path difference is equal to the 
integer number of the wave lengths --- 
the waves interfere positively and produce maximum, in the other extreme case when the path difference is half the integer number of the wave lengths --- 
the waves interfere negatively 
and produce minimum. The result on the screen is a characteristic interference pattern. The second important effect is the strength of the signal. 
This  depends on  the energy deposited on the screen and this energy is proportional to the square of the incoming wave. 
\beq
signal \sim (wave_1 +wave_2)^2
\eeq
\begin{figure}[htb]
\centerline{%
\includegraphics*[width=5cm]{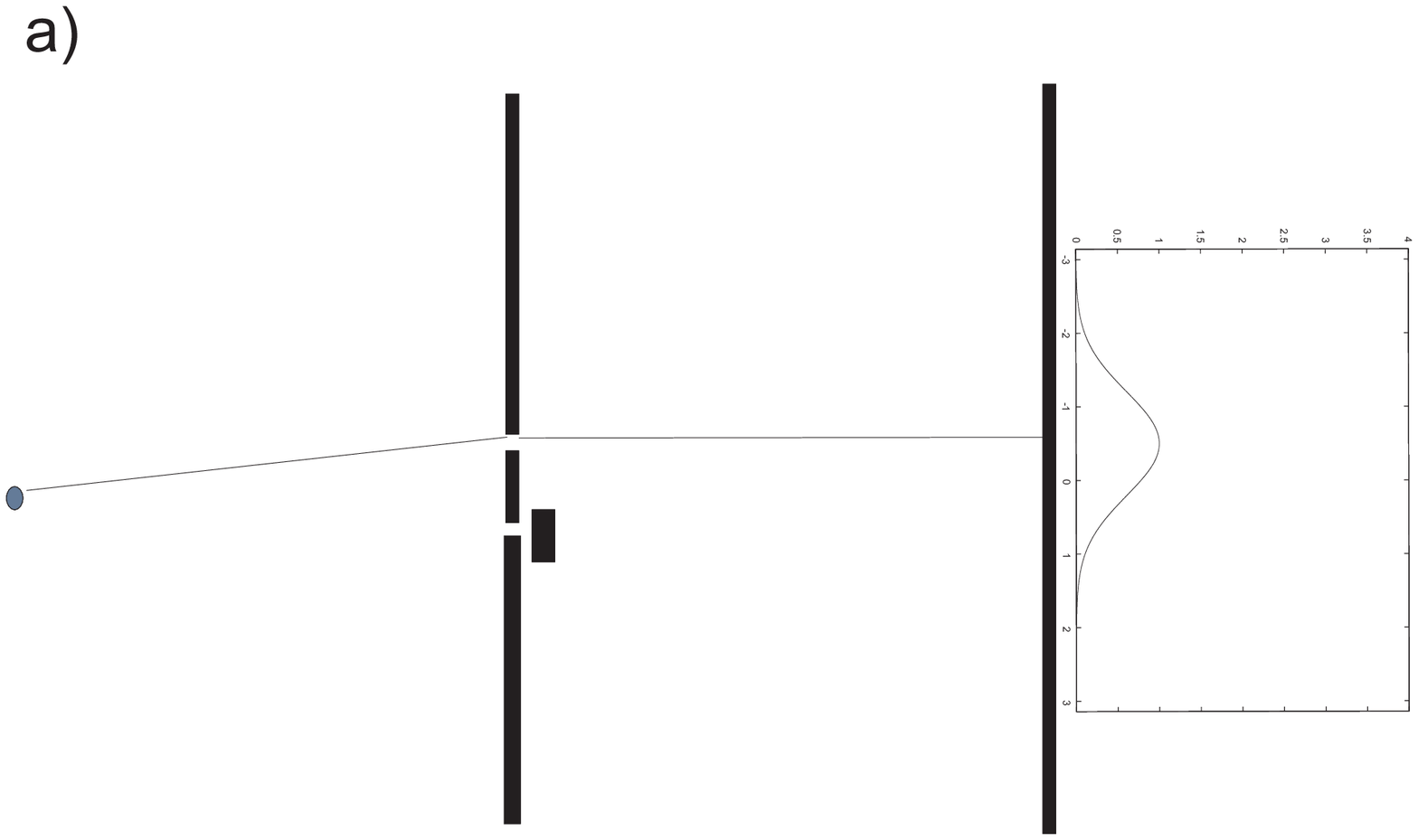}}
\vspace{0.5mm}
\centerline{%
\includegraphics*[width=5cm]{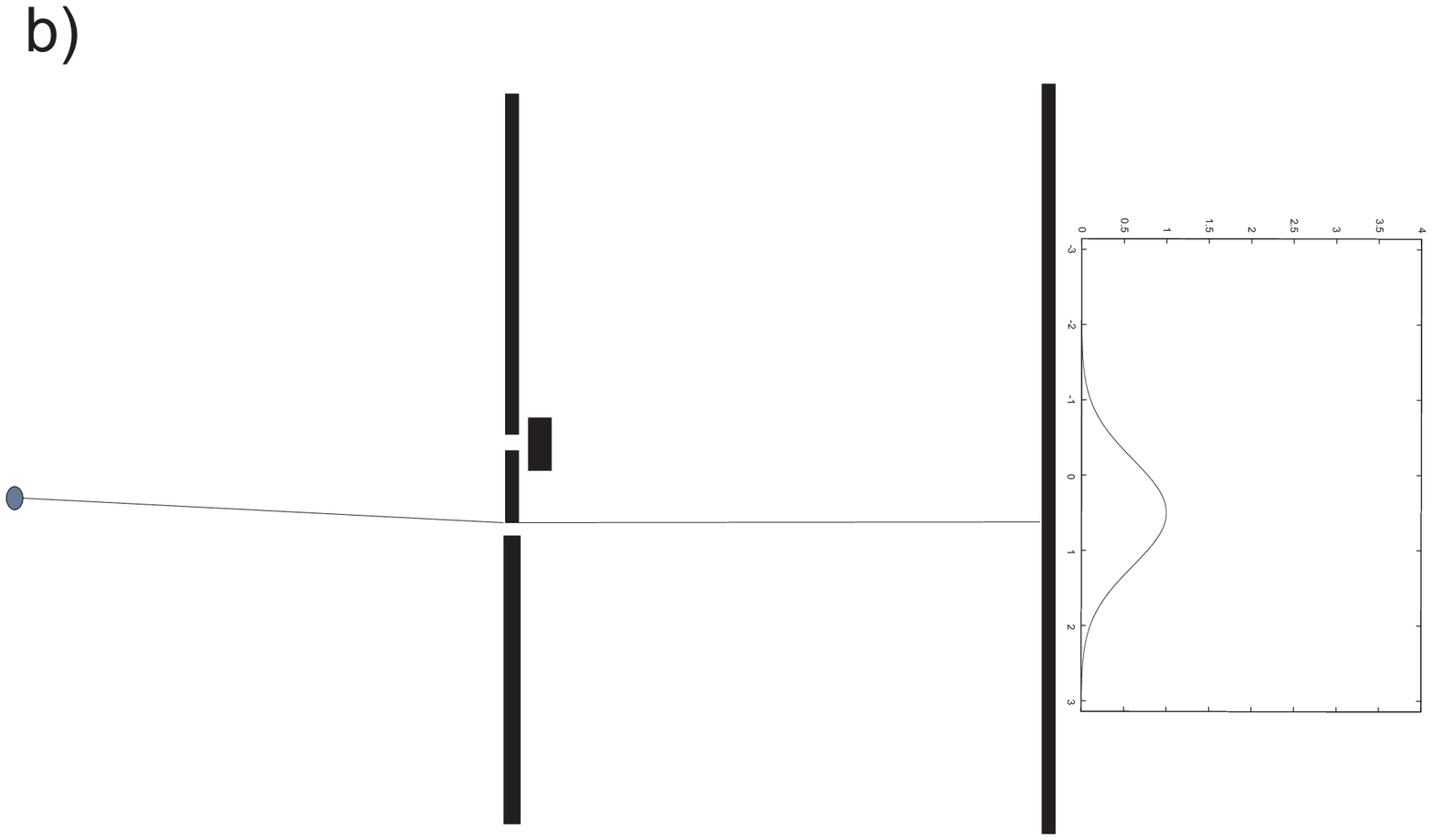}}
\vspace{0.5mm}
\centerline{%
\includegraphics*[width=5cm]{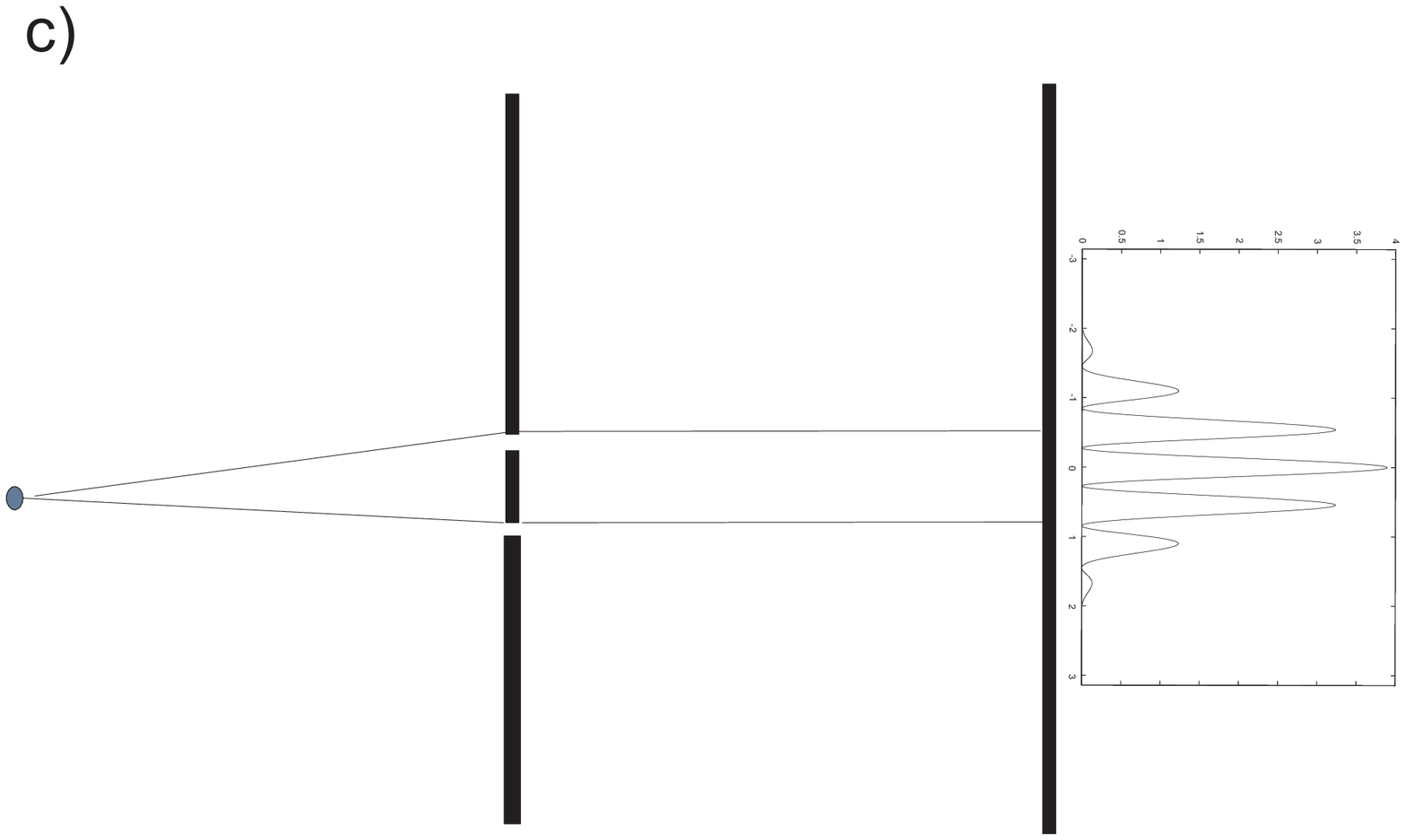}}
\caption{Schematic setup of the two slit interference experiment.}
\end{figure} 

Therefore, when two equal waves meet "in phase" the effective wave is twice as high but the signal strength is 4 times stronger. 
Finally one should stress
an important detail of the experimental setup which makes the observation possible. The interference pattern of light waves is extended in space due to the 
large distance $L$ between the slits and screen as compared to the slit separation $d$ --- approximately by a factor of
$L/d$ . This is why we do not need the accuracy of the order of the wave length $\lambda$ but rather $\lambda L/d$. Choosing properly
the slit separation $d$ and the distance 'slits--target' $L$ we are able to see interference taking place on a nanometric scale 
(light) magnified to a distance in centimetres on the screen.

To summarize the description of double slit experiment, the interference pattern 
following from adding two different waves and "abnormal" amplification of signal at the maximum are the pronounced features of wave phenomena. 

3. The original double slit experiment was repeated more than a~century later in its new version contributing crucially to our understanding of quantum mechanics. 
In 1909 G.I. Taylor used single photons as signal source \cite{Taylor}. Analogical experiments with "quantum waves" 
of matter were also performed with electrons \cite{Jonsson} in 1961, with single electrons \cite{Tonomura} in 1989, with neutrons \cite{Zeilinger} in 1988, with helium atoms \cite{Mlynek} in 
1991 and  with fulleren (C60) molecules \cite{Arndt} in 1999. This series of experiments differs substantially from the "classical wave" phenomenon. In the quantum version 
what we control directly is the 
source of photons or massive particles and the effect on the target (screen). Emitting the single photon or electron we are unable to answer the question 
(without additional measurement) which 
slit has been used by the travelling quantum object. The measurement occurs at the target where the photon/electron appears 
as a nearly pointlike signal. It is only after gathering substantial statistics that these single particles build up
an interference pattern. This pattern  is 
produced by the quantum mechanical probability amplitude which travels between the source and target. The cited experiments support the double, wave-particle  
nature of emitted objects and demonstrate quantum interference 
phenomena at scales as large as fulleren molecules.

4. The double slit interference experiment sketched above has turned out to be a powerful tool in the search for wave phenomena for more than two centuries. 
The aim of this note is to propose an 
analogical experiment for the  brain. We do not attempt to join the speculations on the origins of wave phenomena,  on whether  they are 
classical or quantum-like.  
Instead we would like to concentrate on the experimental feasibility of such experiments. In general, we should like to investigate  the spatial and/or temporal structure 
of effective 
signals resulting from two different sources. 
The main unknown which determines the experimental setup is the scale (in distance and in time) at which the phenomenon manifests itself. Therefore, one is 
forced to perform tests at levels ranging from intercellular (nm) distances, trough single neuron ($\mu m$) to multi neuron (mm) distances and corresponding time scales. 
These scales determine the signal source and the target. 

There are two general setups leading to possible  interference effects.  
The first one is exactly copied from the Young experiment, we call it "spatial" setup. A single source (or two correlated/coherent  sources) emits 
the signal which is carried to the target via two distinct ways. Before hitting the target both signal carriers mix in some overlapping volume 
where they possibly interfere. The target is spatially large enough  that our detectors distinguish details within it (the space resolution of 
our detector should be much smaller than the area reached by the signal). As in the Young experiment, due to different distances covered by the two signal carriers,
the effective signal will show enhanced and suppressed interference bands as we move across the target. Moreover, the strength of the enhanced signal is higher than the simple sum of two single signals. 
{\it Nota bene} to check this, we should be able to send single signal via each single path as well (corresponding to the Young setup with one slit covered).

The second setup uses  time difference rather than path difference, we call it therefore the "temporal" setup. It assumes two sources which can emit signals at 
different (in general), well controlled  instants. The detector in the target, where the effective signal is measured, can be simpler than in the spatial setup, 
it can be just a single measuring unit. The possible 
interference effect will be observed by shifting the relative time of emission of the two signals. The characteristic enhancement-suppression pattern emerges here 
as a function of time difference. As compared to the "spatial" case the "temporal" setup looks simpler but requires more precision. 
The resolution should be high enough so that we are able to distinguish details at times comparable to the wave period. In the "spatial" case 
the interference pattern can be extended in space due to the magnification factor mentioned above.

Both presented setups can be refined or combined to make the search as complete as possible. This is especially important due to our ignorance of  the scale 
(wavelength, frequency) of the phenomenon. The possible quantum-like behaviour requires even more attention --- systematic built up of statistics.

5. How can we look for interference effects in brain? We suggest a few examples at various scales, the list is by no means complete. Let us start with difficulties:
the main ones seem to be connected with the coherence of the sources and the choice of the target. The signal sources can be relatively easily defined, they
 can be external (sensory, vision, sound, ...) 
or internal (direct stimulation of brain cells), the control over their relative correlation is very important. 
The brain as the target is much more  complicated than a  photographic plate, luminescent 
screen or electronic detector. Therefore the correct preparation of the target and the removal, as much as possible, of side effects will be a real challenge. 

One of the best 
systems ready for both "spatial" and "temporal" experiments is the sensory system of the rat. The signal sources here are the rat's vibrissa and the target --- 
the barrel cortex. To facilitate the detection of interference pattern the signal sources should be preferably of equal strength. 
In order to perform the "spatial" experiment one needs a multi-electrode detector which allows for spatial resolution of effective signals. 
The scale at which the effect is searched can be chosen by the magnitude and separation of electrodes, and ranges from single neurons to multi neuron systems. 
The "temporal" experiment seems slightly simpler, it can be performed with single electrode but the time (in particular time shift) control must be very accurate. 
Performing the experiment \it in vivo \rm   one is able to experiment with the brain at rest or in an active state. 
Similar setups can be arranged with the visionary system. The signal sources can be 2 light signals separated in space and/or time, 
their coherence is relatively easy to achieve. Again the right choice of the target 
 and multi-electrode detector would be crucial.

Going down with scale one can test the signal transmission inside the neuron. Although this process seems to be understood at the electrochemical level
 some brave hypotheses \cite{Penrose} 
may be verified in this system. Coherence of the signal emission and transmission could be the main problems here.
Increasing the scale one may use such techniques as EEG, MEG or high precision MRI 
to experiment at multi neuronal systems. These devices allow for a wide choice of sources and targets, their resolution is however much lower --- at present
 in the millimetre (and second) range. The  suggestions   presented above are very much  first guess proposals. But even with this set of experiments
 we are able to cover quite a broad spectrum of wavelengths and frequencies.

6. Let us illustrate the above consideration with a simple model. Assume the signal carrier to be a wave packet of the form (Fig.2 upper left)

\beq
f_1(u) = exp \left(-\frac{u^2}{a^2} \right) cos \left( 2\pi \frac{u}{\lambda} \right) \, .
\eeq
According to our previous remarks the signal detected from such a wave is proportional to the square of the wave (Fig. 2 upper right)
\beq
S_1 \sim \big[ f_1(u) \big]^2
\eeq
and the total strength of such signal is proportional to the integral over the whole signal
\beq
S_1^{\rm int} \sim \int \limits^\infty_{-\infty} S_1 du \, .
\eeq
Let us add the second carrier of the same shape but shifted by a given $\Delta u$  The resulting signal is 
\beq
S_{1+2} \sim \big[ f_1(u) + f_2(u+\Delta u) \big]^2
\eeq
The above formulae are illustrated in Fig. 2 for a few selected shifts $\Delta u$. One can see the interference at work, for some shifts the 
effective signal nearly disappears.

There are several integral measures of relative influence of two signals on each other.  In Fig. 3 we present one of them --- 
 the facilitation index $FI$ defined as the ratio of the integrated effective signal strength over the sum of two separate 
integrated signals
\beq
FI = \frac{S_{1+2}^{\rm int}}{S_1^{int}+S_2^{\rm int}}
\eeq
\begin{figure}
\centerline{%
\epsfig{width=5cm,file=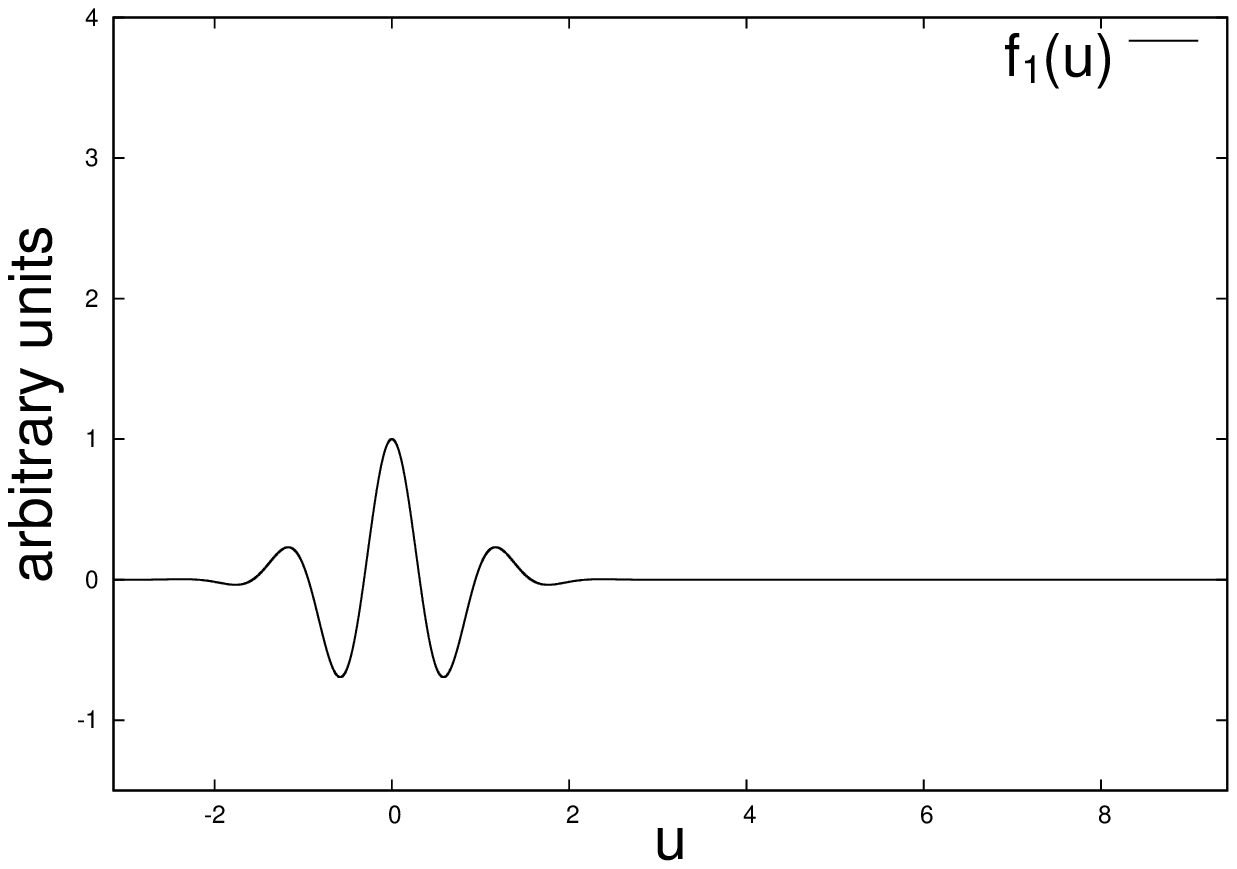}
\epsfig{width=5cm,file=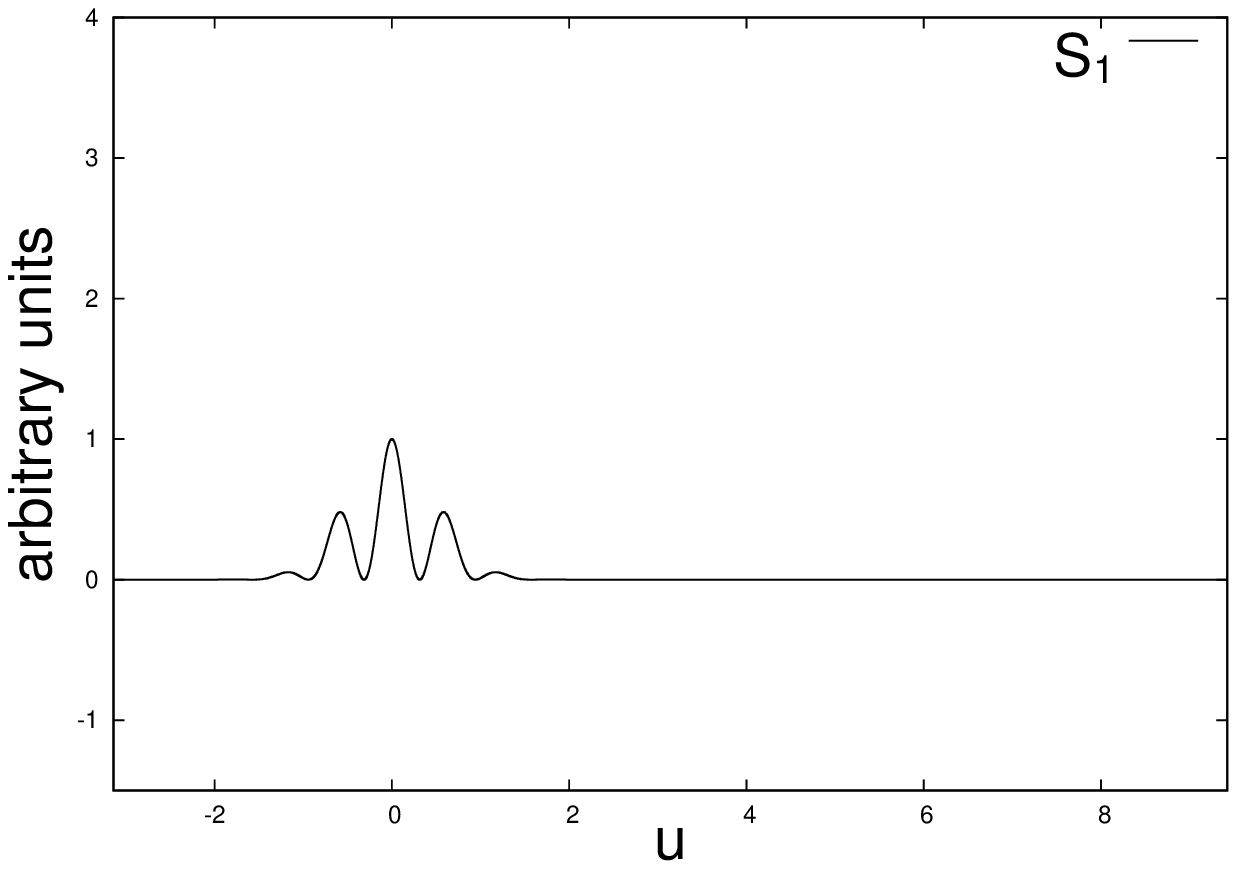}
\put(-270,80){(a)}
}
\centerline{%
\epsfig{width=5cm,file=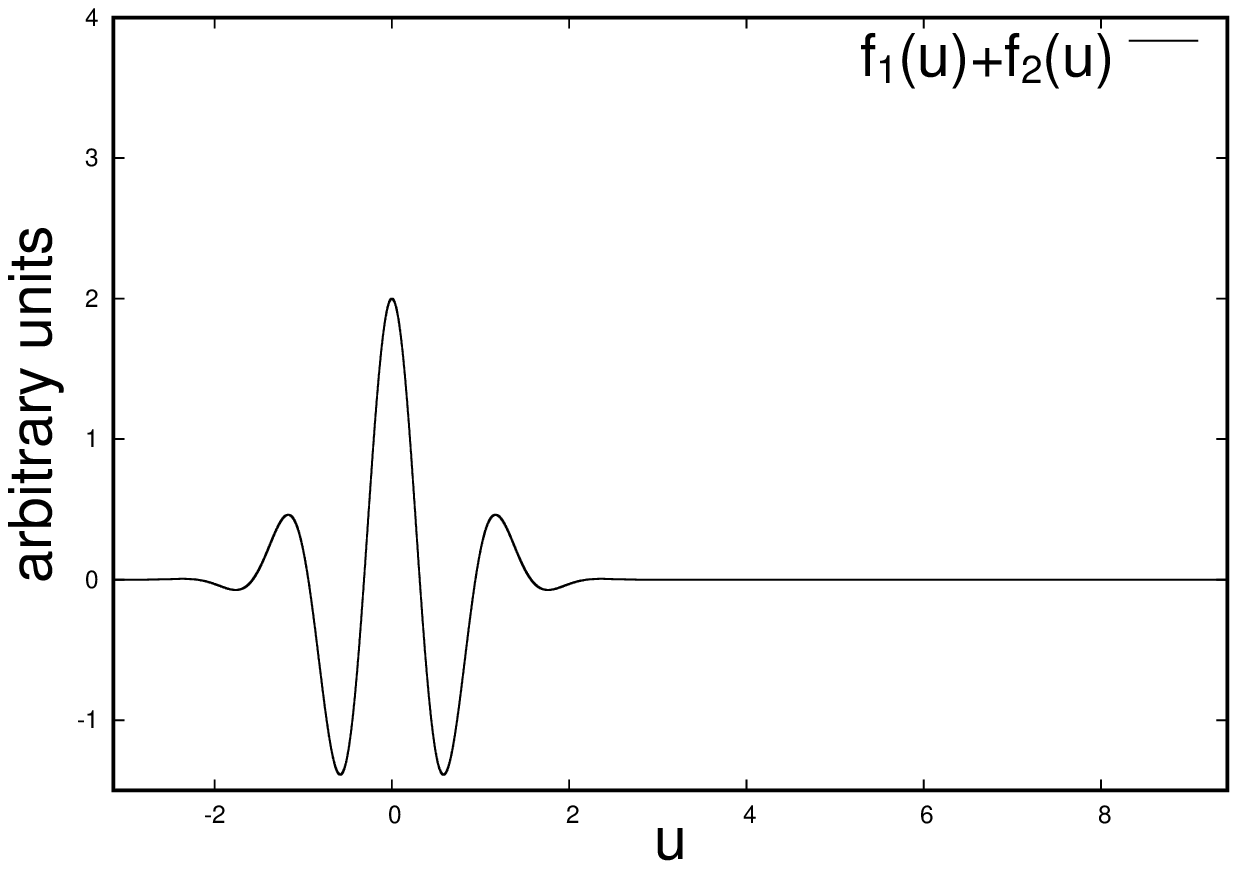}
\epsfig{width=5cm,file=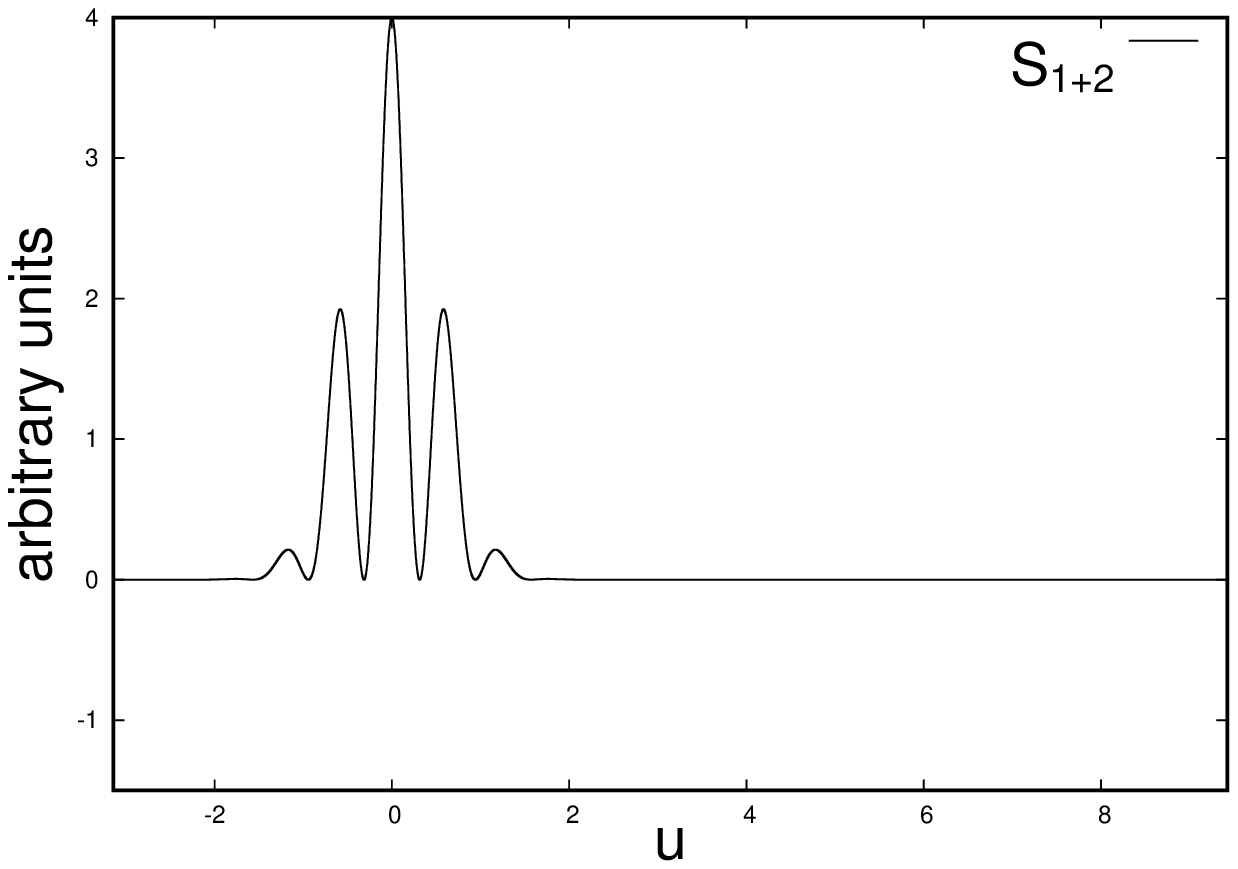}
\put(-270,80){(b)}
}
\vspace{0.5mm}
\centerline{%
\epsfig{width=5cm,file=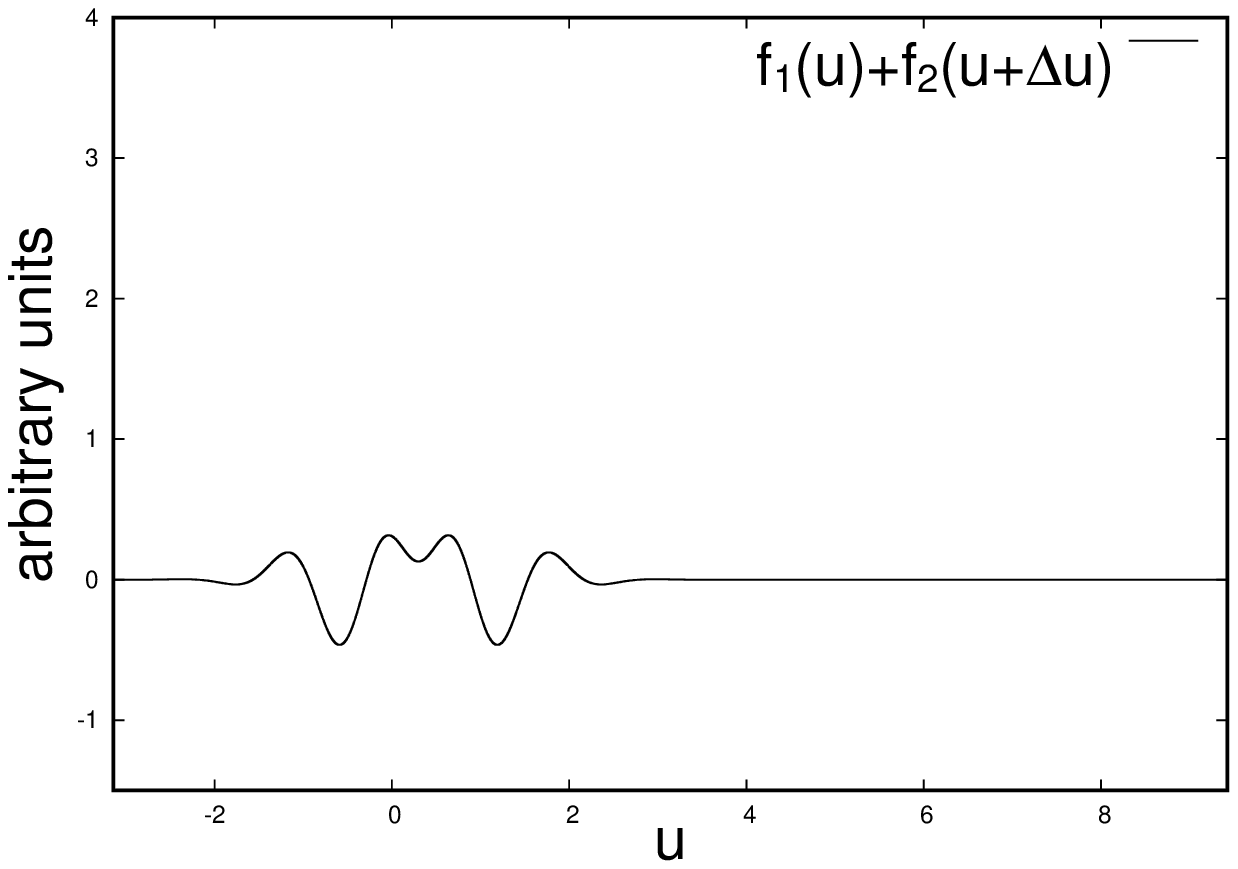}
\epsfig{width=5cm,file=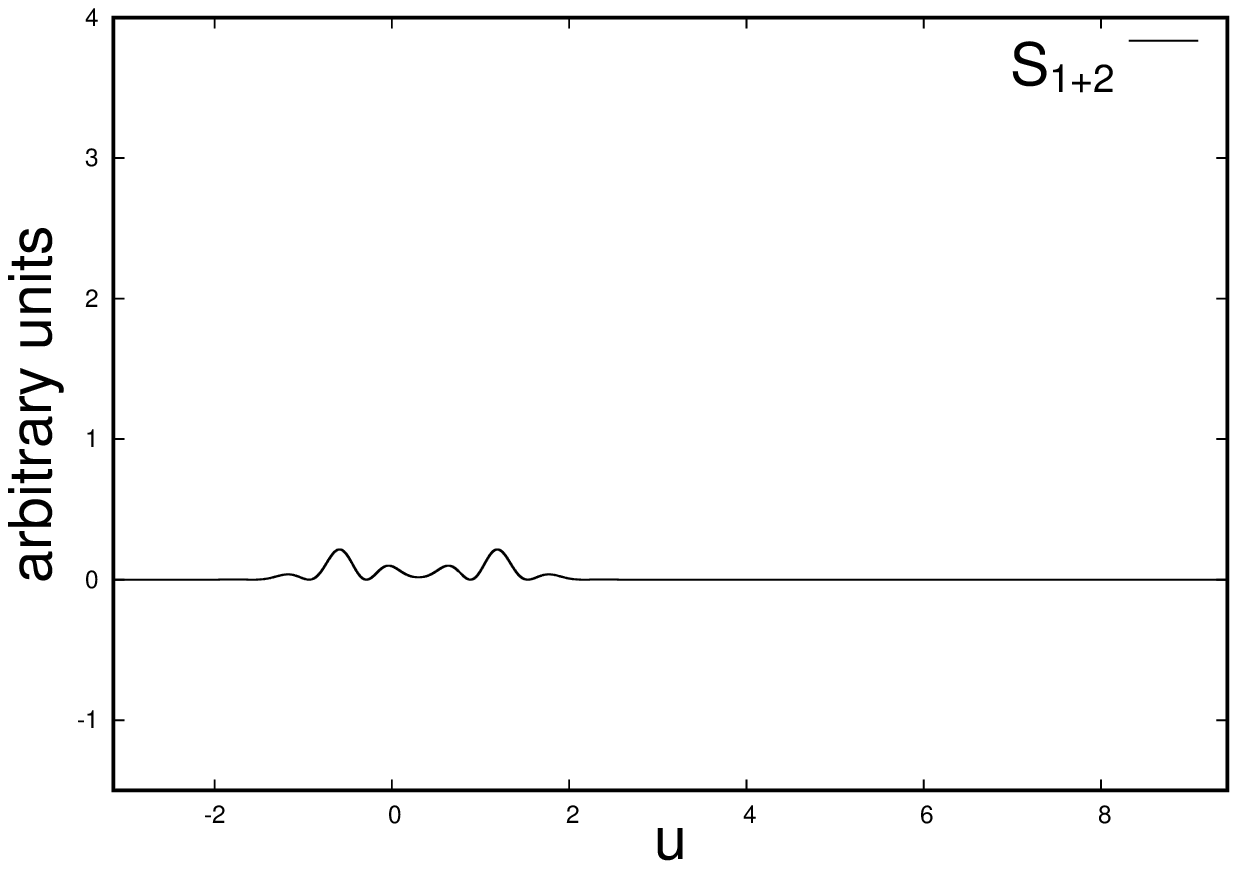}
\put(-270,80){(c)}
}
\vspace{0.5mm}
\centerline{%
\epsfig{width=5cm,file=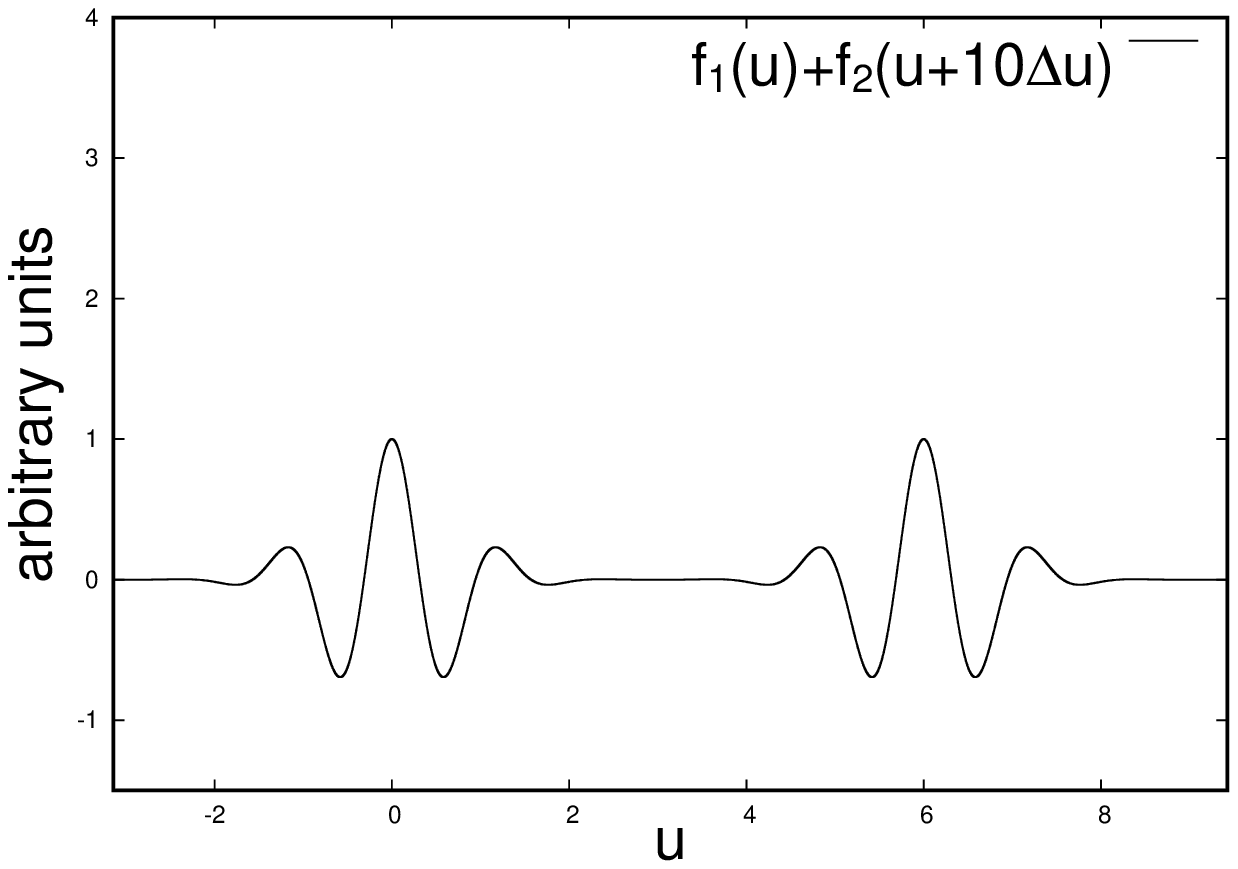}
\epsfig{width=5cm,file=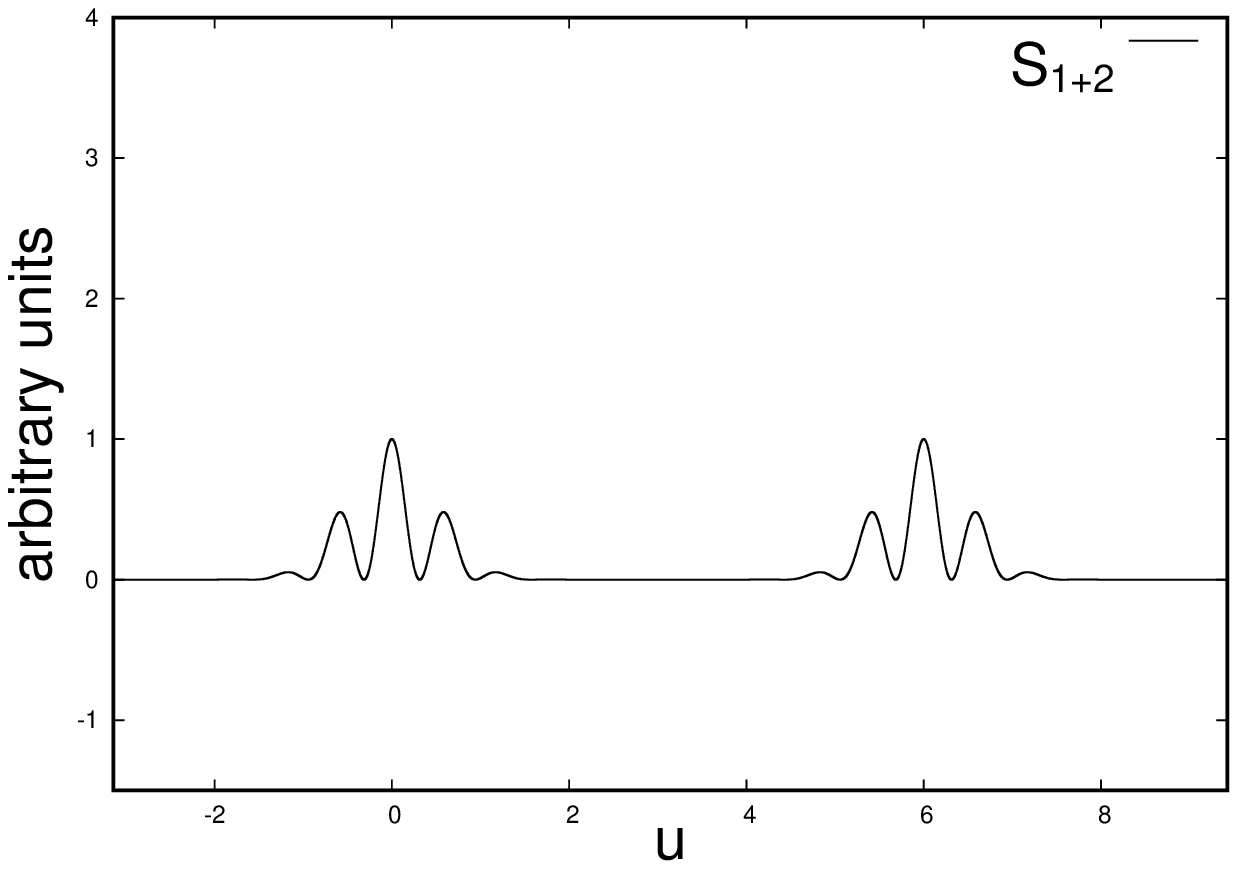}
\put(-270,80){(d)}
}
\caption{(a) Example wave packet and signal. (b)-(d) Interference of two waves and two signals 
for  reletive shifts: (b) $0$, (c) $\Delta u$ and (d) $10 \Delta u$.}
\vspace{-3mm}
\end{figure} 
\begin{figure}
\centerline{%
\includegraphics*[height=4cm, width=8cm]{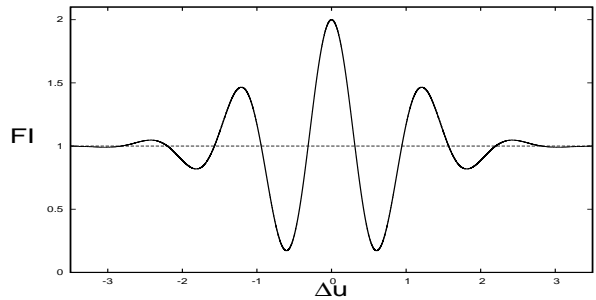}}
\caption{The facilitation index FI as a function of relative shift $\Delta u$.}
\end{figure}
In any case where the signals do not influence each other, this index should be constant and equal to 1. In our example
it shows a characteristic oscillatory behaviour of FI and supralinear enhancement, The ratio FI is thus sensitive to the most prominent features of interference. 
One should note that the variable $u$ (and $\Delta u$) 
does not necessarily mean the spatial dimension. This is so in the original Young ("spatial") setup, 
in the case of the "temporal" setup these variables are  
rather the time and the time shift, respectively.

7. To summarize, this note advocates a systematic search for wave effects in the brain. Following the spectacular double slit interference 
experiment, which has been successful 
so many times and in so many systems, we point out the most characteristic features of the interference pattern. 
 Without knowing the scale of possible  effects it is reasonable to test as many  experimental systems as possible where
  multi stimulus sources and spatio-temporaly extended targets are at our disposal. 
Since we are dealing with a very complex system, the removal of side effects so that we are as close as possible to the original 
two slit experiment will be probably the hardest task.
However, in the case of a positive result  --- the discovery of the wave nature of information processing will certainly be 
 a breakthrough in our understanding of the brain.

8. The author would like to thank Marian H. Lewandowski and Marcin Szwed for discussions.


\end{document}